\documentclass[11pt]{article}

\usepackage[margin=1in]{geometry}
\usepackage{graphicx}
\usepackage[pdftex,colorlinks=true,linkcolor=blue,citecolor=blue,urlcolor=black]{hyperref}
\usepackage{amsmath, amsthm, amssymb,amsfonts,bbm}
\usepackage{subfigure}
\usepackage{comment}
\usepackage{soul}
\usepackage{url}
\usepackage{pdflscape}
\usepackage[ruled,lined,linesnumbered]{algorithm2e}

\newcommand{\bR}{\mathbb{R}}

\newcommand{\bC}{\mathbb{C}}
\newcommand{\bZ}{\mathbb{Z}}

\newcommand{\bL}{\mathbb{L}}

\newcommand{\cL}{\mathcal{L}}
\newcommand{\cM}{\mathcal{M}}
\newcommand{\cA}{\mathcal{A}}
\newcommand{\cS}{\mathcal{S}}
\newcommand{\cE}{\mathcal{E}}

\newcommand{\ket}[1]{| #1 \rangle}
\newcommand{\bra}[1]{\langle #1|}

\newcommand{\proj}[1]{| #1 \rangle \langle #1 |}

\newcommand{\avr}[1]{\left \langle#1 \right \rangle}
\newcommand{\1}{\mathbbm{1}}
\newcommand{\Sp}{\,\,\,\,\,\,}

\newcommand{\tr}[1]{{\rm tr}\left[{#1}\right]}

\DeclareMathOperator{\Var}{Var}

\newcommand{\half}{\frac{1}{2}}

\newcommand{\be}{\begin{equation}}
\newcommand{\ee}{\end{equation}}
\newcommand{\bea}{\begin{eqnarray}}
\newcommand{\eea}{\end{eqnarray}}
\newcommand{\bes}{\begin{equation*}}
\newcommand{\ees}{\end{equation*}}
\newcommand{\beas}{\begin{eqnarray*}}
\newcommand{\eeas}{\end{eqnarray*}}

\makeatletter
\newtheorem*{rep@theorem}{\rep@title}
\newcommand{\newreptheorem}[2]{%
\newenvironment{rep#1}[1]{%
 \def\rep@title{#2 \ref{##1} (restated)}%
 \begin{rep@theorem}}%
 {\end{rep@theorem}}}
\makeatother

\newtheorem{thm}{Theorem}
\newtheorem*{thm*}{Theorem}
\newtheorem{cor}[thm]{Corollary}

\newtheorem*{lem*}{Lemma}

\newreptheorem{thm}{Theorem}
\newreptheorem{lem}{Lemma}

\begin{document}

\title{Non-commutative Nash inequalities}
\author{Michael Kastoryano$^1$ and Kristan Temme$^2$\\[8pt]
{\small $^1$ NBIA, Niels Bohr Institute, University of Copenhagen, 2100 Copenhagen, DK}\\
{\small $^2$ Institute for Quantum Information and Matter, California Institute of Technology,}\\
{\small Pasadena CA 91125, USA}
}

\date{\today}

\maketitle

\begin{abstract}
A set of functional inequalities -- called \textit{Nash inequalities} -- are introduced and analyzed in the context of quantum Markov process mixing. The basic theory of Nash inequalities is extended to the setting of non-commutative $\bL_p$ spaces, where their relationship to Poincar\' e and log-Sobolev inequalities are fleshed out. We  prove Nash inequalities for a number of unital reversible semigroups. 
\end{abstract}


Bounding the mixing times of quantum channels is becoming an increasingly important topic in quantum information sciences, with applications ranging from self-correcting quantum memories \cite{temme2014} to Gibbs sampling \cite{kastoryano2014} to quantum Shannon theory \cite{muller2013,kastoryano2012} and many-body dynamics \cite{cubitt2013,kastoryano2013r}. The main task in the mixing time analysis of quantum dynamical semigroups is to find the least time after which the following bound holds: 
$||e^{t\cL^*}(\sigma)-\rho||_1\leq \epsilon$, for any initial state $\sigma$ and some desired precision $\epsilon$. Here $\rho$ is the stationary state of the semigroup $S_t^*\equiv e^{t\cL^*}$\footnote{$S_t$ will act on observables (the Heisenberg picture), and $S_t^*$ on states (the Schr\"odinger picture). The conjugation is with respect to the Hilbert-Schmidt inner product}. Operationally, the mixing time corresponds to the time after which all  information in $\sigma$ is erased by the noise process $S_t$.  Upon optimization over all input states, the trace norm bound can be rewritten as a 

\be \sup_{||g||_\infty\leq 1}||S_t(g)-\rho(g)||_\infty \equiv\xi(t)\leq \epsilon,\ee
where $\rho(f)\equiv \tr{\rho f}\1$. In general, assuming that the generator of the semigroup satisfies detailed balance (see Eqn. (\ref{eqn:db})), a generic bound available on $\xi(t)$ is $\xi(t)\leq \sqrt{||\rho^{-1}||} e^{-t \lambda}$, where $\lambda$ is the spectral gap of $\cL$ (see Eqn. (\ref{def:gap})). However, $||\rho^{-1}||$ scales as the size of the phase space (dimension of the operator algebra), and often overshoots by quite a bit on the actual mixing time. We show that if the generator $\cL$ of the semigroup satisfies a \textit{non-commutative Nash inequality}
\be ||f-\rho(f)||_{2,\rho}^{2+4/\nu}\leq C\avr{f,-\cL(f)}_\rho||f||_{1,\rho}^{4/\nu} \label{eqn:NashI},\ee
with some positive constants $C,\nu\in\bR^+$ for all observables $f$, then the following improvements on the mixing time can be obtained:

\bea ||S_t-\rho||_{\infty\rightarrow\infty}&\leq& ||S_{t_1}-\rho||_{2\rightarrow\infty,\rho}||S_{t_2}-\rho||_{2\rightarrow2,\rho}\\
&=&||S_{t_1}-\rho||_{2\rightarrow\infty,\rho}e^{-t_2 \lambda},\eea
where $t_1+t_2=t$, and the last line follows from the definition of the spectral gap (Eqn. \ref{spectGapVar}).
The norms appearing in Eqn. (\ref{eqn:NashI}) are non-commutative $\bL_p$ norms \cite{pisier2003,zegarlinski2000,kastoryano2013}  with respect to the stationary state $\rho$ of $S^*_t$. 
If the semigroup satisfies a non-commutative Nash inequality, by Theorem \ref{thm:Nash1}, we get
\be ||S_t-\rho||_{2\rightarrow\infty,\rho}\leq \left(\frac{\nu C}{4t}\right)^{\nu/4}.\ee
Then, taking the supremum over $t_1+t_2=t$, yields 
\be ||S_t-\rho||_{\infty\rightarrow\infty}\leq 2e^{-\lambda (t-\nu C/4)},\label{mixingbd}\ee
whenever $t\geq \nu C/4$. Thus if a Nash inequality can be established with good constants $\nu$ and $C$, then extremely tight bounds on the mixing time of the semigroup can be obtained, which are often optimal when the underlying graph structure of the phase space is highly constrained but regular. To the knowledge of the authors, Nash inequalities only appear to be very powerful for  single particle problems on local regular graphs.  In general, obtaining good estimates on the constants $\nu,C$ can be very challenging.  

In this paper, we extend the theory of Nash inequalities to the non-commutative setting. We show that two different types of Nash inequalities lead to close to optimal mixing time bounds. A lower bound on the Log-Sobolev constant is proved in terms of the constants $\nu,C$ in the Nash inequality, as well as strong bounds on the entire spectrum of the generator $\cL$. Finally, as applications we prove Nash inequalities for the depolarizing semigroup, for unital qubit semigroups and for a random walk on the ring.

\section{Preliminaries}

We denote the set of $d\times d$ complex matrices $\cM_d$ and the subset of Hermitian matrices $\cA_d=\{X\in\cM_d,X=X^\dag\}$, as well as the subset of positive definite matrices $\cA^+_d=\{X \in \cA_d, X > 0\}$. The set of states will be denoted $\cS_d=\{X \in \cA_d,X\geq0,\tr{X}=1\}$, and the full rank states will be analogously denoted $\cS_d^+$. 
Observables will always be represented by lower case Latin letters ($f,g\in \cA_d$), and states by Greek letters ($\rho,\sigma \in \cS_d$). 

Non-commutative Nash inequalities are defined with respect to finite dimensional non-commutative $\bL_p$ spaces \cite{pisier2003}. The non-commutative $\bL_p$ spaces are characterized by a norm and an inner product, which for any $f,g\in\cA_d$  and some $\rho\in\cS_d^+$, are defined as
\be \label{eqn:lpnorms}
\| f \|_{p,\rho}^p =\tr{\rho^{1/2p}f\rho^{1/2p}}, ~~~~~~~{\rm and}~~~~~~~~~\avr{f,g}_\rho=\tr{\rho^{1/2}f\rho^{1/2}g}
\ee
We will also make extensive use of the $\bL_p$ variance which is defined as 
\be\Var_\rho(g) =\avr{f-\rho(f),f-\rho(f)}_\rho, \ee 
where $\rho(f)=\tr{\rho f}\1$ is the uniform projection onto the expectation of $f$ with respect to $\rho$.  

\bigskip 

Throughout this paper, the time evolution of an observable ($f_t\in\cA_d$) will be described by one-parameter semigroups of completely positive trace preserving maps (cpt-maps), whose generator (Liouvillian) can always be written in standard \textit{Lindblad form} 
 \be
\partial_t f_t =  \cL(f_t) \equiv i[H,f_t] + \sum_i L^\dag_i f_t L_i - \frac{1}{2}\{L^\dagger_i L_i , f_t\}_+,
\ee
where $H\in\cA_d$ is the system Hamiltonian, and accounts for the coherent dynamics in the system, and the $L_i\in\cM_d$ are Lindblad operators describing dissipation on the system. Note that we have written the dynamics on operators (Heisenberg picture) for convenience. We denote the dual of $\cL$, with respect to the Hilbert-Schmidt inner product, by $\cL^*$ which amounts to the evolution of states, i.e. the Schr\"odinger picture.  The trace preserving condition ensures that $\cL(\1)=0$. If in addition $\cL^*(\1)= 0$, then the dynamics are said to be unital. A Liouvillian $\cL$  is said to be \textit{primitive} if it has a unique full-rank stationary state. As the framework of non-commutative $\bL_p$ spaces depends on a full rank reference state, which will most often be the stationary state of a some dissipative dynamics, we will almost exclusively consider primitive Liouvillians. The reference state $\rho$ should always be clear from the context, and will almost always be the unique full rank stationary state of $S_t^*\equiv e^{t\cL^*}$.

 We say a Liouvillian $\cL:\cM_d\rightarrow\cM_d$ satisfies \textit{detailed balance} (or is \textit{reversible}) with respect to the state $\rho\in\cS_d^+$, if for any $f,g\in\cA_d$
\be\label{eqn:db} \avr{f,\cL(g)}_\rho=\avr{\cL(f),g}_\rho\ee
The class of reversible generators has a number of particularly nice properties. The one most often exploited is that if $\cL$ satisfies detailed balance with respect to some $\rho\in\cS_d^+$, then $\rho$ is a stationary state of $\cL$.  Furthermore, the detailed balance condition ensures that the generator is Hermitian with respect to the weighted inner product $\avr{f,g}_\rho$, which ensures that $\cL$ has a real spectrum. 

We will frequently need the \textit{Dirichlet form} of $\cL$: $\cE(f)=-\avr{f,\cL(f)}_\rho$, which can be used to characterize the \textit{spectral gap} of a primitive reversible generator $\cL$ as \cite{chi2}:

\bea \label{spectGapVar}
	\lambda  = \min_{g \in \cA_d}  \frac{\cE(g)}{\Var_\rho(g)}.\label{def:gap}
\eea


\section{Nash Inequalities}

The results presented in this paper aim to develop a useful theory of Nash inequalities for quantum dynamical semigroups. Most of our arguments follow the classical works of Diaconis, Saloff-Coste \cite{diaconis1996,saloff1997}. In some cases, the classical and quantum proofs are almost identical. We try as much as possible to point out what aspects of the quantum theory differ from the classical theory. 

The cornerstone of the theory of Nash inequalities for mixing time analysis is a partial equivalence between so-called \textit{ultracontractivity} of the semigroup and a Nash inequality. This connection is reminiscent of the relation between hypercontractivity and the Log-Sobolev inequalities \cite{diaconis1996,zegarlinski2000,carbone2008,kastoryano2013}. The following theorem sketches out one direction of the implication: 

\begin{thm}\label{thm:Nash1}[Nash I]
Let $\cL:\cM_d\rightarrow\cM_d$ be a primitive reversible Liouvillian. If for all $f\in\cA_d$, $\cL$ satisfies a {\rm type I Nash inequality}  
\be \Var_\rho(f)^{1+2/\nu}\leq C\cE(f)||f||_{1,\rho}^{4/\nu}\label{eqn:Nash1}\ee
for some positive constants $\nu,C\in\bR^+$, then 
\be ||S_t-\rho||_{1\rightarrow 2}\leq \left(\frac{\nu C}{4t}\right)^{\nu/4},\ee
for all $t\geq 0$.
\end{thm}
\proof{
Set $u(t):= ||S_t(f)-\rho(f)||_{2,\rho}^2\equiv\Var(S_t(f))$, then letting $||f||_{1,\rho}=1$,
Eqn. (\ref{eqn:Nash1}) can be rewritten as 
\be u(t)^{1+2/\nu}\leq -\frac{C}{2}\dot{u}(t),\ee
which is equivalent to $\dot{v}(t)\geq 1$, where $v(t)=\frac{C \nu}{4}u(t)^{-2/\nu}$. Solving for $v$ and recasting the result in terms of $u$ yields
\be u(t)\leq  \left(\frac{\nu C}{4t}\right)^{\nu/2}.\ee
Taking the supremum over all functions $f$ with $||f||_{1,\rho}=1$ yields 
\be ||S_t-\rho||_{1\rightarrow 2,\rho}\leq\left(\frac{\nu C}{4t}\right)^{\nu/4}.\label{eqn:mixing}\ee
\qed}

Note that by duality of $\bL_p$ norms and reversibility of the semigroup, one immediately also gets a bound on the $\bL_{2\rightarrow \infty}(\rho)$ norm for the semigroup, which was necessary to obtain the mixing time bound of Eqn. (\ref{mixingbd}). Furthermore, invoking the Riez-Thorin interpolation theorem (see eg. Ref. \cite{beigi2013sandwiched}), the bound can be extended to give:
\be 
||S_t-\rho||_{1\rightarrow \infty,\rho}\leq\left(\frac{\nu C}{4t}\right)^{\nu/2}.
\ee
Some semigroups are not ultracontractive for all times, but relax more rapidly for short times than a Poincar\' e inequality would suggest. In these cases, a different type of Nash inequality is more useful, that zeros out the short time convergences of the semigroup. 

\begin{thm}\label{thm:Nash2}[Nash II]
Let $\cL:\cM_d\rightarrow\cM_d$ be a primitive reversible Liouvillian. If for all $f\in\cA_d$, $\cL$ satisfies a {\rm type II Nash inequality}   
\be ||f||_{2,\rho}^{2+4/\nu}\leq C(\cE(f)+\frac{1}{T}||f||_{2,\rho}^2)||f||_{1,\rho}^{4/\nu} \label{eqn:Nash2}\ee
for some positive constants $\nu,C,T\in\bR^+$, then for all $t\leq T$, 
\be ||S_t||_{1\rightarrow 2,\rho}\leq e^{t/T}\left(\frac{\nu C}{4t}\right)^{\nu/4}.\label{eqn:ultra2}\ee
\end{thm}

\proof{The proof is very similar to that of Theorem \ref{thm:Nash1}. Again we fix $f$ satisfying $||f||_{1,\rho}=1$, but now define 
\be u(t)=e^{-2t/T}||S_t(f)||^2_{2,\rho}.\ee
Then, 
\be \dot{u}(t)=-2 e^{-2t/T}(\cE(S_t(f))+\frac{1}{T}||S_t(f)||^2_{2,\rho})\ee
Eqn. (\ref{eqn:Nash2}) now reads
\be u^{1+2/\nu}(t)\leq -\frac{C}{2}\dot{u}(t)\ee
Using the same argument as in the proof of Theorem \ref{thm:Nash1}, we get
\be u(t)\leq \left(\frac{\nu C}{4t}\right)^{\nu/2},\ee
which implies 

\be ||S_t||_{1\rightarrow 2,\rho}\leq e^{t/T}\left(\frac{\nu C}{4 t}\right)^{\nu/4}.\ee\qed}

The type II Nash inequality says something strong about the mixing at short times, but does not provide any information about the asymptotic behavior. However, the ultracontractive bound in Eqn. (\ref{eqn:ultra2}) is sufficient to obtain the improved mixing time bound of Eqn. (\ref{mixingbd}), provided that $T\geq \nu/(4\lambda)$, where $\lambda$ is the spectral gap of $\cL$. 

Ultracontractivity (i.e. $||S_t||_{2 \rightarrow \infty,\rho}\leq1$) is related to Nash inequalities in the sense that one is the infinitesimal formulation of the other, as we will see in the next theorem. This behavior is very reminiscent of the relationship between Hypercontractivity and Log-Sobolev inequalities \cite{zegarlinski2000,kastoryano2013} or the contraction of the $||S_t||_{2 \rightarrow 2, \rho}$  norm and the  Poincare inequality, Eqn. (\ref{spectGapVar}). One can show (c.f. Theorem \ref{thm:Nash1} and \ref{thm:Nash2} ) that Ultracontractivity not only follows from a Nash inequality, but also that the converse is true. It is this fact, that ensures that the mixing time bounds obtained from Nash inequalities are in fact close to optimal for good constants $(C,\nu,T)$, since the contractive bound on the $\bL_{2\rightarrow \infty}(\rho)$ norm eliminates the dependence on any generic pre factor. 


\begin{thm}\label{thm:converse}
Let $\cL:\cM_d\rightarrow\cM_d$ be a primitive reversible Liouvillian, and suppose that the semigroup $S_t\equiv e^{t\cL}$ satisfies 
\be ||S_t-\rho||_{1\rightarrow 2,\rho}\leq \left(\frac{C}{t}\right)^{\nu/4},\label{eqn:ultraI}\ee for all $t\geq0$ and some positive constants $\nu,C\in\bR^+$, then
\be \Var_\rho(f)^{1+2/\nu}\leq C'\cE(f)||f||_{1,\rho}^{4/\nu},\label{eqn:NashI}\ee
for all $f\in\cA_d$, where $C'=2^{2+4/\nu}C$. Furthermore, if for all $t\leq T$, 
\be ||S_t||_{1\rightarrow 2,\rho}\leq \left(\frac{C}{t}\right)^{\nu/4},\label{eqn:ultraII}\ee
for some positive constants $\nu,C,T\in\bR^+$, then 
\be ||f||_{2,\rho}^{2+4/\nu}\leq C'\left(\cE(f)+\frac{1}{T}||f||_{2,\rho}^2\right)||f||_{1,\rho}^{4/\nu},\label{eqn:NashII}\ee
for all $f\in\cA_d$, where again $C'=2^{2+4/\nu}C$.
\end{thm}

\proof{
We will only prove that Eqn. (\ref{eqn:ultraI}) implies Eqn. (\ref{eqn:NashI}). The proof that Eqn. (\ref{eqn:ultraII}) implies Eqn. (\ref{eqn:NashII}) is very similar, and a classical version of it can be found in Ref. \cite{saloff1997}. 
 
Fix $f$ with $||f||_{1,\rho}=1$ and write, for $0\leq t\leq T$, 
\bea \Var_\rho(f)&=& \Var_\rho(f_t)-\int_0^tds \frac{d}{ds}\Var_\rho(f_s)\\
&=&\Var_\rho(f_t)+2\int_0^tds \cE(f_s)\\
&\leq& (C/t)^{\nu/2}+2t\cE(f).\label{eqn:rhsVar}\eea
The inequality follows from the ultracontractivity assumption Eqn. (\ref{eqn:ultraI}), and from the observation that for all $t\geq0$ and $f\in\cA$, $\cE(f)\geq\cE(f_t)$. To see the latter, note that by primitivity, $\sqrt{\cL}$ is well defined and commutes with $\cL$, and thus also with $S_t$. Then, by reversibility, 
\bea \cE(f_t)&=& \avr{f_t,-\cL(f_t)}_\rho\\
&=&||\sqrt{-\cL}e^{t\cL}(f)||^2_{2,\rho}\\
&=&||e^{t\cL}\sqrt{-\cL}(f)||^2_{2,\rho}\\
&\leq&||\sqrt{-\cL}(f)||^2_{2,\rho}=\cE(f),\eea
where the last inequality follows form contractivity of the $\bL_{2\rightarrow2}(\rho)$ norm. 
Solving for the minimum of the right hand side of Eqn. (\ref{eqn:rhsVar}), we recover the Nash I inequality
\be \Var_\rho(f)^{2+4/\nu}\leq B\cE(f)||f||_{1,\rho}^{4/\nu},\ee
with 
\be B=2C(1+2/\nu)(1+d/2)^{2/\nu}\leq 2^{2+4/\nu}C\ee\qed}

As will be evident in the example section, the converse theorem is more useful for Nash II inequalities, as it is often much simpler to obtain a bound of the form Eqn. (\ref{eqn:ultraII}) for some range of $t$ up to a cutoff $T$, than for all real $t$. 

\subsection{Lower bound on the Log-Sobolev constant}

Nash inequalities are functional inequalities involving $\bL_p$ norms and Dirichlet forms. Many other functional inequalities (Poincar\'e, Sobolev, Log-Sobolev) have been extensively studied, both in the classical and in the quantum setting. Here we relate the constants appearing in the Nash inequalities with the Log-Sobolev constant, as defined in Refs. \cite{kastoryano2013,zegarlinski2000,temme2014h}.  

\begin{thm}\label{LS-BndFromNash2}
Let $\cL:\cM_d\rightarrow\cM_d$ be the generator of a primitive reversible semigroup. If for all $f\in\cA_d$, $\cL$ satisfies
\be \Var_\rho(f)^{1+2/\nu}\leq C\cE(f)||f||_{1,\rho}^{4/\nu} \ee
for some positive constants $\nu,C\in\bR^+$, then the Log-Sobolev constant $\alpha_2$ is bounded below by
\be \alpha_2\geq \frac{2}{\nu C}\label{eqn:NashLS1}\ee
If $\cL$ satisfies 
\be ||f||_{2,\rho}^{2(1+2/\nu)}\leq C\left(\cE(f)+\frac{1}{T}||f||_{2,\rho}^2\right)||f||_{1,\rho}^{4/\nu},\ee
for some positive constants $\nu,C,T\in\bR^+$, with $\nu C/4\leq T$ and has spectral gap $\lambda$, then the Log-Sobolev constant $\alpha_2$ is bounded below by
\be \alpha_2\geq \frac{\lambda}{2(1+\lambda t_0+\nu/4 \log(\frac{\nu C}{4 t_0}))}\ee
for any $0<t_0\leq T$.
\end{thm}

\proof{
From the bound in Eqn. (\ref{mixingbd}) we get that for $t_0=\nu C/4\leq T$, 
\be  ||S_t||_{2\rightarrow\infty,\rho}\leq 1\ee
Then by Theorem 5 in Ref. \cite{temme2014h}, we simply get
\be \alpha_2 \geq  \frac{\lambda}{2(1+\lambda t_0+\nu/4 \log(\frac{\nu C}{4 t_0}))}.\ee
The proof of Eqn. (\ref{eqn:NashLS1}) is essentially identical, but using a variant of Theorem 5 in Ref. \cite{temme2014h} that can be found (in classical form) in Ref. \cite{saloff1997} Theorem 2.2.13.\qed}

The Log-Sobolev constant $\alpha_2$ lower bounds the spectral gap $\lambda$ for primitive reversible channels (see Ref. \cite{kastoryano2013}), which completes the chain of inequalities between Nash, Log-Sobolev and Poincar\'e constants. \\

Finally, we point out that the Nash inequalities are implied by non-commutative Sobolev inequalities. Indeed, consider the  following non-commutative Sobolev inequalities:

\bea ||f-\rho(f)||^2_{2\nu/(\nu-1),\rho}&\leq& C\cE(f)\label{Sobolev1}\\
||f||^2_{2 \nu/(\nu-2),\rho}&\leq&C\left(\cE(f)+\frac{1}{T}||f||^2_{2,\rho}\right)\label{Sobolev2}.\eea
By H\" olders inequality, 
\be ||f||_2^{2(1+2/\nu)}\leq||f||^2_{2\nu/(\nu-2)}||f||_1^{4/\nu},\ee
and we immediately see that Eqn. (\ref{Sobolev1}) implies a type I Nash inequality, and Eqn. (\ref{Sobolev2}) implies a type II Nash inequality. Unlike in the classical setting, the converse is not known in the non-commutative setting.

\subsection{Bounds on higher eigenvalues}

One particular appeal of the Nash inequalities is that they imply bounds on the entire spectrum (i.e. higher eigenvalues), rather than just on the spectral gap.  Let $\cL$ be a reversible Liouvillian and let $0\leq \lambda_0\leq\lambda_1\leq ...\leq\lambda_{n-1}$ be the eigenvalues of $-\cL$, and define the counting function:

\be N(s)=\# \{i\in\{0,1,...,n-1\}|\lambda_i\leq s\}, ~~~~s\geq 0.\ee
$N$ is a step function with $N(s)=1$ for $0\leq s<\lambda_1$ (for primitive $S_t$). It is easy to relate the counting function $N$ to the eigenvalues of $-\cL$. Define the one parameter function
\be \zeta(t)=\sum_i e^{-t\lambda_i}\ee

Since for $\lambda_i\leq 1/t$ we get $e^{-t\lambda_i}\geq e^{-1}$, it follows that 
\be N(1/t)\leq e\zeta(t)\ee

Thus if we can find an upper bound on $\zeta(t)$, we can bound the counting function.  Here we show that $\zeta(t)$ can be bounded using Nash inequalities. 

\begin{thm}\label{thm:spectralbd}
Let $\cL:\cM_d\rightarrow\cM_d$ be a primitive reversible Liouvillian. If for all $f\in\cA_d$, $\cL$ satisfies
\be \Var_\rho(f)^{1+2/\nu}\leq C\cE(f)||f||_{1,\rho}^{4/\nu} \ee
for some positive constants $\nu,C\in\bR^+$, then the counting function satisfies
\be N(s)\leq 1+e(\nu Cs/2)^{\nu/2}\label{eqn:evalNashI}\ee
for all $s\geq 0$. If the generator satisfies a type II Nash inequality
\be ||f||_{2,\rho}^{2(1+2/\nu)}\leq C\left(\cE(f)+\frac{1}{T}||f||_{2,\rho}^2\right)||f||_{1,\rho}^{4/\nu},\ee
then \be 
N(s)\leq e^3(\nu Cs/2)^{\nu/2}\label{eqn:evalNashII}
\ee

for all $s\geq 1/T$.
\end{thm}

\proof{We will prove the bound for type II Nash inequalities. The proof of the bound for type I is very similar. 
We will use a mapping between quantum channels and matrices, whereby $S(f)=\sum_k E_k f E_k^\dag$ gets mapped to $\hat{S}=\sum_k E_k\otimes \bar{E}_k$. Similarly, matrices $X$ get mapped onto states as $\ket{X}=X\otimes\1\ket{\Omega}$, where $\ket{\Omega}=\sum_j \ket{j,j}$ is proportional to the maximally entangled state.  

Now define the map $\Gamma_\rho(f)=\rho^{1/2}f\rho^{1/2}$, and the \textit{symetrized} semigroup as $\tilde{S}_t=\Gamma_\rho^{1/2}S_t\Gamma_\rho^{-1/2}=\Gamma_\rho^{-1/2}S_t^*\Gamma_\rho^{1/2}$. Since $\tilde{S}_t$ and $S_t$ have the same spectrum, we get 

\bea
\sum_je^{-t\lambda_j}&=& \tr{\hat{\tilde{S}}_t}\\
&=&\sum_j \avr{K_j|\hat{\tilde{S}}_t|K_j}\\
&=& \sum_j \tr{K_j\tilde{S}_t(K_j)}\\
&=&\sum_j \avr{L_j,S_t(L_j)}_\rho\\
&=&\sum_j||S_{t/2}(L_j)||_{2,\rho}^2,
\eea
where $\{\ket{K_j}\}$ form some orthonormal basis, and $L_j=\Gamma_\rho^{-1/2}(K_j)$.
Then, from a type II Nash inequality, we have
\be ||S_{t/2}||^2_{1\rightarrow 2,\rho}\leq e^{2t/T}\left(\frac{\nu C}{2 t}\right)^{\nu/2}\equiv \epsilon(t),\ee
hence,

\bea
 \sum_je^{-t\lambda_j}&\leq& \epsilon(t) \sum_j ||L_j||^2_{1,\rho}\\
&=& \epsilon(t) \sum_j \tr{\rho^{1/2}K_j}^2\\
&=&\epsilon(t) \sum_j   \avr{\sqrt{\rho} |K_j} \avr{K_j|\sqrt{\rho} } = \epsilon(t) \tr{\rho}.
\eea
 
Thus, we get that whenever a type II Nash inequality holds, the eigenvalue counting function satisfies Eqn. (\ref{eqn:evalNashII}) for all $s\geq 1/T$. In order to show the bound in Eqn. (\ref{eqn:evalNashI}), one simply needs to use the  identity $||S_t(h)-\rho||_{2,\rho}^2=||S_t(h)||_{2,\rho}^2-1$.\\
\qed}

Clearly, if $M(s)$ is a continuous and monotonously increasing function such that $N(s)\leq M(s)$, $s\geq1/T$, then 
\be 
\lambda_i\max\{s:N(s)\leq i\}\geq M^{-1}(i+1)
\ee

for all $i>M(1/T)-1$. Hence, we obtain,

\begin{cor}
Let $\cL:\cM_d\rightarrow\cM_d$ be a primitive reversible Liouvillian, which satisfies a type I Nash inequality, then 
\be \lambda_j\geq \frac{2 j^{2/\nu}}{e^{2/\nu}\nu C}.\ee
If $\cL$ satisfies a type II Nash inequality, then
\be  \lambda_j\geq \frac{2 (j+1)^{2/\nu}}{e^{6/\nu}\nu C},\ee
for all $j>e^3\left(\frac{\nu C}{2T}\right)^{\nu/2}-1$. 
\end{cor}

\section{Examples}

We provide a number of examples which illustrate the strengths and the weaknesses of the Nash inequalities. The general intuition about Nash inequalities is that they are good at identifying local uniform mixing for short times, but are not very good at bounding global or long time mixing properties. For this Poincar\'e and Log-Sobolev inequalities are typically better.  As we will see, the only example where Nash inequalities provide the right mixing time estimate is for a one particle on a very regular grid. This is consistent with classical results that have not proven very useful for many body systems, but have provided the best mixing time results for random walks on regular grids or groups (see Ref. \cite{diaconis1996}). 

\subsection{The depolarizing semigroup}

As our first example, we prove a Nash I inequality for the depolarizing semigroup: $\cL(f)=\gamma (\tr{f\rho}\1-f)$. Its fixed point is $\rho$. 
The Nash inequality is  saturated for projectors $f=\ket{\psi}\bra{\psi}$. 

Let $(C \gamma)^{\nu/4}\geq ||\rho^{-1}||^{1/2}$, then 

\be \Var_\rho(f)\leq ||f||^2_{2,\rho}\leq (C \gamma)^{\nu/2}||f||^2_{1,\rho}\ee

Recalling that $\cE(f)=\gamma \Var_\rho(f)$ for the depolarizing semigroup, we recover the Nash I inequality
\be \Var_\rho(f)^{1+2/\nu}\leq C\cE(f)||f||_{1,\rho}^{4/\nu}\ee

Then maximizing $C\nu/4$ subject to $||\rho^{-1}||^{1/2}\leq (C\gamma)^{\nu/2}$ yields a mixing time of 
\be \tau_{\rm mix}=\frac{e}{2} \frac{\log(||\rho^{-1}||)}{\gamma^2}\ee

Clearly, this bound provides no improvement on the one obtained from the spectral gap alone.

\subsection{Unital qubit channels}

Next we consider reversible unital qubit semigroups, and show that the Nash inequalities can be easily obtain as a function of the parameters of the King-Ruskai representation of qubit channels. 
We consider a representation of quantum channels that was first considered in Ref. \cite{king2001}. Any qubit density operator can be represented as
\be \rho=\half(\1 +\vec{w}\cdot \vec{\sigma})\ee
where $\vec{w}\in\bR^3$ and $|\vec{w}|\leq1$ is a real vector and $\vec{\sigma}=(\sigma_X,\sigma_Y,\sigma_Z)$ are the Pauli matrices. Pure states are characterized by the sub-manifold satisfying $|\vec{w}|=1$. A quantum channel $S:\bC^2\rightarrow\bC^2$ acts on the state $\rho$ as
\be S(\rho)=\half(\1+(\vec{s}+\Delta \vec{\omega}))\cdot\vec{\sigma}),\ee
for some vector $\vec{s}$ and matrix $\Delta$. 

It can then be seen that the quantum channel $S$ is unital if and only if $\vec{s}=0$, and is reversible if and only if $\Delta$ is Hermitian. Lets now consider the case of a unital reversible qubit semigroup $S_t:\bC^2\rightarrow\bC^2$. Then the action of the semigroup on a state $\rho$ is $S_t(\rho)=\half(\1+ (\Delta_t \vec{w})\cdot \vec{\sigma})$, where $\Delta_t=e^{t L}$ is now also a semigroup. Denote the eigenvalues of $\Delta_t: (\lambda_1^t=e^{-t l_1},\lambda_2^t=e^{-t l_2},\lambda_3^t=e^{-t l_3})$. Then given that $S_t$ is Hermitian and unital, $\lim_{t\rightarrow\infty}\Delta_t=0$, which implies that $(l_1,l_2,l_3)$ are all real and strictly positive. 

We can now estimate the $\bL_{1\rightarrow2}(\1/2)$ norm of $S_t$:
\bea ||S_t||^2_{1\rightarrow2,\1/2}&=& 2\sup_{\tr{\rho}=1}\tr{\rho S_{2t}(\rho)}\\
&=& \sup_{|\vec{w}|=1}(1+\vec{w}\cdot \Delta_{2t}\vec{w})\\
&=&1+e^{-2tl_{\rm min}}\leq \left(\frac{1}{l_{\rm min}t}\right)^{1/2},\eea
for $t\leq 1/2l_{\rm min}$ where $l_{\rm min}=\min\{l_1,l_2,l_3\}$. Then, from Theorem \ref{thm:converse}, we get the Nash II inequality:
\be ||f||_{2,\rho}^{6}\leq \frac{2^6}{l_{\rm min}}\left(\cE(f)+\frac{1}{2l_{\rm min}}||f||_{2,\rho}^2\right)||f||_{1,\rho}^{4}.\ee

We might be tempted to extend the Nash II inequality  to the case of a tensor power of reversible unital qubit channels. This can be done by invoking the multiplicativity of Shatten $1\rightarrow2$ norms for unital channels, proved in Ref. \cite{king2002,king2014hypercontractivity}. Since for any unital qubit channel $N$, 

\be
 ||S^{\otimes N}||_{1\rightarrow2}=||S||_{1\rightarrow2}^N,
\ee
we get for the unital reversible semigroup described above that

\be \label{eqn:qubitN}
||S_t^{\otimes N}||^2_{1\rightarrow2,\1/2^N}\leq\left(\frac{2}{l_{\rm min}t}\right)^{N/2}.
\ee

Eqn. (\ref{eqn:qubitN}) then immediately leads to a Nash II inequality with constants $\nu=N$, $C'=2^{2+4/N}/l_{\rm min}$ and $T = (2l_{\rm min})^{-1}$. Note that for $N > 2$ the constants will inevitably violate the assumption that $\nu(4 \lambda)^{-1} \leq T$, which is a necessary condition to infer the mixing time bound from the constants in the Nash II inequality in Theorem \ref{thm:Nash2}. This is a simple example of the pathological behavior of Nash inequalities under multiplication.  As a matter of fact the ill behavior under the tensor product operation of Nash Inequalities and in turn Sobolev inequalities was a central motivation to introduce Log-Sobolev inequalities \cite{gross1975logarithmic}. It is known that this semi-group has in fact a system size independent Log-Sobolev constant $\alpha_2$ \cite{king2014hypercontractivity,montanaro2008quantum,temme2014}.

\subsection{Single particle diffusion and coherent hopping}

As our last example we consider a finite chain of length $N$ with periodic boundary conditions. We assume, that a single particle can hop dissipatively in both directions with equal rates. Moreover, we allow for additional coherent transport with a single particle Hamiltonian $H$. This model is defined on the Hilbert space $\bC^{N+1}$ and can be viewed as the one-particle restriction of the quantum mechanical version of the symmetric exclusion process as considered in \cite{eisler2011crossover}. We will see that the particular choice of the Hamiltonian $H$ does not affect the convergence rate so that the equilibration bound holds for a large class of coherent hopping models. 

The Lindblad operators that mediate the diffusive hopping are defined by $L_m = \ket{m}\bra{ m \oplus_N 1 }$ for all $m = 0,\ldots,N$.  We denote by $\oplus_N$ the addition modulo $N$ in $\bZ_N$. Moreover, we allow for the coherent transport by any Hamiltonian $H$ on $\bC^{N+1}$, such as for examples $H = \sum_m L_m+ L_m^\dag$. The semi-group can be written as

\be\label{hop-gen}
 \cL_C(f)= i [H,f] +\cL(f), \Sp \mbox{where} \Sp \cL(f) = \half \sum_{m=0}^N \; [L_m,[f,L_m^\dag]] + [L_m^\dag,[f,L_m]].
\ee

The stationary state of $\cL_C$ is the maximally mixed state $\rho = (N+1)^{-1}\1$. For any hermitian $f \in \cM_{N+1}(\bC)$ we have that $-i\tr{f^\dag [H,f]} = 0$, which implies the bound $\cE_C(f) = -(N+1)^{-1}\tr{f^\dag \cL _C(f)}$ by the form $\cE(f) = -(N+1)^{-1}\tr{f^\dag \cL (f)}$. That way, we have $\cE_C(f) \leq \cE(f)$ and the Nash II inequality Eqn. (\ref{eqn:NashII}) automatically holds for $\cL_C$ with the same constants $(C,\nu, T)$ as for $\cL$.

We now follow the same approach as in the previous example  and bound the $\bL_{1 \rightarrow 2}(\1/(N+1))$ norm of the semi-group $S_t = \exp(t\cL)$ that is generated by the Liouvillian $\cL$ in Eqn. (\ref{hop-gen}). We note (see Ref. \cite{watrous}) that the supremum in the definition of $|| S_t ||_{1\rightarrow 2} = \sup_f  ||S_t ||_2 \; || f ||^{-1}_1$ for Shatten norms $|| f ||^p_p  = \tr{|f|^p}$ is always attained for rank one projectors $f = \proj{\psi}$. Hence, we obtain for the $\bL_{1 \rightarrow 2}(\1/(N+1))$ norm: 

\bea\label{hop-normBND}
	\|S_t\|_{1 \rightarrow 2, \1/(N+1)}^2 &=& (N+1) \sup_{ \avr{\psi | \psi} = 1 } \bra{\psi}S_{2t}(\proj{\psi}) \ket{\psi}\\
	&=&(N+1)  \sup_{ \avr{\psi | \psi} = 1 }\sum_{m,l=0}^Ne^{-2t\lambda_{ml}}|\bra{\psi}v_{ml}\ket{\psi}|^2\label{eqn:chain1},
\eea
where $\lambda_{ml}$ are the eigenvalues of $-\cL$ and $v_{ml}$ are its (right) eigenvectors. 
From basic Fourier analysis, the eigenvectors and eigenvalues can all be evaluated analytically, yielding

\bea
\begin{array}{lll}
 0 \leq m \leq N: \Sp    & \lambda_{m0} = \;  -2(1 - \cos(\frac{2 \pi m}{N+1}))  \Sp & v_{m0}   =  \frac{1}{\sqrt{N+1}} \sum_{k=0}^N \; e^{i\frac{2\pi m k }{N+1}} \; \proj{k} \vspace{0.3cm}\\
 l\neq 0    \Sp   & \lambda_{ml} = \; -2	\Sp &  v_{ml}   = \ket{m}\bra{m\oplus l}.
\end{array}
\eea
Then, Eqn. (\ref{eqn:chain1}) can be evaluated explicitly
\bea \|S_t\|_{1 \rightarrow 2, \1/(N+1)}^2 &=&(N+1)\sum_{m=0}^N(e^{-2t\lambda_{m0}}|\bra{\psi}v_{m0}\ket{\psi}|^2+\sum_{l=1}^Ne^{-t\lambda_{ml}}|\bra{\psi}v_{ml}\ket{\psi}|^2)\\
&\leq&\sum_{m=0}^N(e^{-2t\lambda_{m0}}+(N+1)\sum_{l=1}^Ne^{-2t\lambda_{ml}}),
\eea
by observing that $\sum_m|\bra{\psi}v_{ml}\ket{\psi}|^2\leq 1$ for all $l\neq0$, which follows from simple linear algebra. We get the expression for $l=0$ that $\sum_{k=0}^N \exp\left(i\frac{2\pi m k }{N+1}\right)|\psi_k|^2 \leq \avr{\psi|\psi}$, where  $\ket{\psi} := \sum_i \psi_i \ket{i}$. We may now bound the $\bL_{1\rightarrow 2}(\1/(N+1))$ norm of the semigroup as:
\be\label{dimsum}
\|S_t\|_{1 \rightarrow 2, \1/(N+1)}^2 \leq  1 + \sum_{m=1}^N e^{-2t(1- \cos(\frac{2\pi m}{(N+1)})} + (N+1)^2 e^{-4t} ,
\ee
since we consider the supremum over normalized states $\ket{\psi}$. By using a trick from Ref. \cite{diaconis1996}, we may bound the exponential of the cosine by a quadratic function and bound the sum in Eqn. (\ref{dimsum}) by a Gaussian integral which then yields
\be
	\sum_{m=1}^N e^{-2t(1- \cos(\frac{2 \pi m}{(N+1)})}\leq 2\sum_{m=1}^N e^{-4t \frac{m^2}{(N+1)^2}} \leq 2e^{-\frac{4t}{(N+1)^2}}\left(1 + \sqrt{\frac{(N+1)^2}{4t}}\right).
\ee
Finally, we observe that for $t \leq T = (N+1)^2/16$ this expression can be bounded by the polynomially decaying function 
\be
\|S_t\|_{1 \rightarrow 2, \1/(N+1)}^2 \leq \left( 1 + 2e^{-\frac{4t}{(N+1)^2}}\left(1 + \sqrt{\frac{(N+1)^2}{4t}}\right) + (N+1)^2 e^{-4t} \right) \leq \sqrt{\frac{8(N+1)^2}{t}}.
\ee

The constants in the ultracontractive bound can be read off directly: $C = 8(N+1)^2$ and $\nu =1$ for $ T =  (N+1)^2/16$. Due to the comparison argument with $\cE(f)_C \geq \cE(f)$ and by virtue of Theorem \ref{thm:converse}, we can state the Nash II inequality for the semigroup generated by $\cL_C$,
\be 
||f||_{2,\1/(N+1)}^{6}\leq 2^{10}(N+1)^2\left(\cE(f)+\frac{16}{(N+1)^2}||f||_{2,\1/(N+1)}^2\right)||f||_{1,\1/(N+1)}^{4}.
\ee
We can therefore infer a mixing time bound that behaves as 
\be
\tau_{\rm mix}= {\cal O}((N+1)^2),
\ee
which gives the exact mixing time, since a matching lower bound can be obtained by inputing a rank one trial function. By comparison, the  spectral gap bound estimate yields $|\lambda_1| \geq (N+1)^{-2}$ that would scales together with the pre factor $\log(||\rho^{-1}||)$ as $\tau_{\rm mix}= {\cal O}((N+1)^2 \log(N+1))$ and therefore the Nash inequality  completely removes the pre factor contribution.
 
\subsection*{Acknowledgements}

M.J.K. was supported by the Carlsbergfond and the Villum foundation. 
K.T. was supported by the Institute for Quantum Information and Matter, a NSF Physics Frontiers Center with support of the Gordon and Betty Moore Foundation (Grants No. PHY-0803371 and PHY-1125565).


\bibliographystyle{plain}
\bibliography{nashbib}

\end{document}